\documentclass[aps,prc,twocolumn, superscriptaddress,showkeys,nofootinbib]{revtex4-1}  

\usepackage[utf8]{inputenc}

\usepackage{graphicx,color}
\usepackage{amsmath,amssymb,amsfonts}
\usepackage{hyperref}

\usepackage{xspace}	% Include xspace

\def  \first      {\mbox{$\textit{v}_{1}$}\xspace}
\def  \second     {\mbox{$\textit{v}_{2}$}\xspace}
\def  \third      {\mbox{$\textit{v}_{3}$}\xspace}

\def \cnm2   {\mbox{$\mathrm{C}_{nm}\lbrace 2 \rbrace$ }\xspace}
\def \cnm4   {\mbox{$\mathrm{C}_{nm}\lbrace 4 \rbrace$ }\xspace}

\def \sc23  {\mbox{$\mathrm{SC}(2,3)$ }\xspace}
\def \sc24  {\mbox{$\mathrm{SC}(2,4)$ }\xspace}

\def \nsc23 {\mbox{$\mathrm{NSC}(2,3)$}\xspace}
\def \nsc24 {\mbox{$\mathrm{NSC}(2,4)$}\xspace}

\newcommand{\psirp}{\Psi^{\rm  RP}}

\begin{document}
\title{Measuring differential flow angle fluctuations in relativistic nuclear collisions
}
%--------------------------------------------------------------------------------------------------------------------------------------------
\medskip
%--------------------------------------------------------------------------------------------------------------------------------------------
%--------------------------------------------------------------------------------------------------------------------------------------------
\author{Niseem~Magdy} 
\email{niseemm@gmail.com}
\affiliation{Department of Chemistry, State University of New York, Stony Brook, New York 11794, USA}
%--------------------------------------------------------------------------------------------------------------------------------------------

%--------------------------------------------------------------------------------------------------------------------------------------------
%\date{\today}
%--------------------------------------------------------------------------------------------------------------------------------------------
\begin{abstract}
%--------------------------------------------------------------------------------------------------------------------------------------------
Recently studies of the differential nature of the flow angle fluctuations, known as event plane angular decorrelation, indicated that measurements that assume a common symmetry plane may need to consider the flow angle fluctuations effect. 
{  Using the HIJING and AMPT models, it is shown that the flow angle fluctuations measurements, obtained with the two- and three-subevents correlation method, could have significant non-flow effects associated with long- and short-range non-flow correlations.
The current study demonstrates the four-subevents cumulant methods' ability to reduce the non-flow effects.
It is further argued that the measurements using the four-subevents correlation method can be used to provide accurate quantification of the differential flow angle fluctuations.}

%I demonstrate that the non-flow effects are mostly suppressed using the four-subevents cumulant methods by requiring azimuthal correlation between the four pseudorapidity ranges.  
%
%I argue that the measurements using the four-subevents correlation method can be used to provide accurate quantification of the differential flow angle fluctuations. 
%----------------------------------------------------------------------
\end{abstract}
%----------------------------------------------------------------------
%\pacs{25.75.-q, 25.75.Gz, 25.75.Ld}% PACS, the Physics and Astronomy
                             % Classification Scheme.
%----------------------------------------------------------------------
%Use showkeys class option if keyword
%----------------------------------------------------------------------
\maketitle
%----------------------------------------------------------------------
%\linenumbers
%----------------------------------------------------------------------

\section{Introduction}
%----------------------------------------------------------------------
%Introduction
%----------------------------------------------------------------------

%----------------------------------------------------------------------
Studies at the Large Hadron Collider (LHC) and the Relativistic Heavy Ion Collider (RHIC) aim to create and study the properties of the deconfined nuclear medium named quark-gluon plasma (QGP)~\cite{Shuryak:1978ij,Shuryak:1980tp,Muller:2012zq}.
In non-central collisions, the medium created by the participant (nucleons that experience at least one collision) interactions will experience a hydrodynamic expansion with an initial geometry defined by the participant distribution, with event-by-event fluctuations.
%In a hydrodynamic evolution, the pressure gradients of the created medium  transform the spatial anisotropies of the initial state into final-state momentum anisotropies.
%
In a hydrodynamic evolution, the created medium's pressure gradients transform the initial state's spatial anisotropies into final-state momentum anisotropies.
Accordingly, the azimuthal distributions of the particles created in the  collisions can be analyzed with a Fourier expansion~\cite{Voloshin:1994mz,Poskanzer:1998yz}:
%----------------------------------------------------------------------
%
\begin{eqnarray}
\dfrac{dN}{d\varphi} &\propto & 1 + \sum_{n=1}^{\infty} 2 v_{n} \cos\left[  n(\varphi -\psirp) \right] ,
\label{equ:Fourier_expansion}   
\end{eqnarray}
where $\varphi$ illustrates the azimuthal angle of a the particle, $v_n$ is the n$^{th}$ order Fourier coefficients, and $\psirp$ is the direction of the reaction plane. The n$^{th}$ Fourier coefficients can be provided as:
\begin{equation}
v_n = \langle \cos [n(\varphi -\Psi_n)] \rangle \,,  
\label{eq:expansionRP}
\end{equation}
where $\Psi_n$ is the flow angles event plan, \first is called directed flow, \second is the elliptic flow, and \third is the triangular flow, etc.
%----------------------------------------------------------------------

%----------------------------------------------------------------------
While $v_n$ fluctuations and the effect of their variance on different methods for measuring $v_n$ have been studied extensively~\cite{Voloshin:2008dg,Heinz:2013th}, flow angle fluctuations (i.e., flow decorrelations) have recently found attention~\cite{Qiu:2011iv,Teaney:2010vd,Gardim:2011xv,Teaney:2012ke,Jia:2012ma,Jia:2012sa,Qiu:2012uy,Ollitrault:2012cm,Gardim:2012im} both theoretically and experimentally.
The flow angles $\Psi_n$ fluctuations (decorrelations) depend on transverse momentum ($p_T$)~\cite{Gardim:2012im,Gardim:2017ruc,Zhao:2017yhj,Bozek:2018nne,Barbosa:2021ccw} and pseudorapidity ($\eta$)~\cite{Bozek:2010vz,Jia:2014ysa,Pang:2014pxa,Pang:2015zrq,CMS:2015xmx,Bozek:2017qir,Cimerman:2021gwf} leading to additional fluctuations between the integrated event plane and the $p_T$ or $\eta$ differential event plane.
Such correlations can lead to a factorization breaking (i.e., the two-particle angular correlations do not factorize into a product of single-particle flow coefficients). The  $p_T$ and $\eta$ dependence of the flow angles $\Psi_n$ fluctuations are studied using two-particle correlations~\cite{Bozek:2010vz,Jia:2014ysa} and, recently, four-particle correlations~\cite{Nielsen:2020Wo,ATLAS:2017rij,ALICE:2022smy,Bozek:2021mov}. 
The longitudinal and transverse momentum flow angle fluctuations give essential insight into the nature of the initial state event-by-event fluctuations of the heavy-ion collisions. Therefore, studying the differential flow angle fluctuations can reflect its impact on the measurements that assume a common symmetry plane. 
%----------------------------------------------------------------------

%----------------------------------------------------------------------
The flow angle fluctuations are measured using the two- and four-particle correlations~\cite{Bozek:2010vz,Jia:2014ysa,Bozek:2021mov}. The  two- and four-particle correlations are  susceptible to long-range non-flow correlations (e.g., jets in a dijet event) and short-range non-flow correlations (e.g., resonance decays, Bose-Einstein correlation, and fragments of individual jets).
Such non-flow correlations usually involve a few particles from one or more $\eta$ regions. The non-flow effects are usually reduced by correlating particles from two or more subevents divided in pseudorapidity.
Therefore, a more detailed study of the influence of non-flow effects on these observables is required before interpreting the experimental measurements. Event generators such as the HIJING model~\cite{Wang:1991hta,Gyulassy:1994ew}, which contain only non-flow correlations, are an ideal testing ground for estimating the influence of non-flow on the four-particle correlations, which are part of the focus of this paper.
%----------------------------------------------------------------------

%----------------------------------------------------------------------

{ 
The present study investigates the non-flow effects on the flow angles $\Psi_n$ fluctuations measurements and its $p_T$ dependence using the HIJING~\cite{Wang:1991hta,Gyulassy:1994ew} model. In addition, this study used the A Multi-Phase Transport (AMPT)~\cite{Lin:2004en} model to investigate the proposed observables, which have been shown to have minimal non-flow effects, on measurements of the $p_T$ dependence of the flow angles $\Psi_n$ fluctuations. The models and analysis method used are given in~\ref{sec:2}. The results and discussion are presented in section~\ref{sec:3}. The conclusions of this work are summarized in section~\ref{sec:4}.
}

%In section I summarize the conclusions and discuss the proposed observables implications.
%--------------------------------------------------------------------

%--------------------------------------------------------------------
%--------------------------------------------------------------------
\section{Methodology} \label{sec:2}
%--------------------------------------------------------------------
\subsection{Models}
%--------------------------------------------------------------------

The present analysis is performed with events simulated by HIJING~\cite{Wang:1991hta,Gyulassy:1994ew} and the AMPT (v2.26t9b)~\cite{Lin:2004en} models for Au+Au (Pb+Pb) collisions at  $\sqrt{s_{NN}}$ = 200(5020)~GeV.
In both models, charged particles with $0.2 < p_T < 5.0$ GeV/$c$, and $|\eta| < 5.0$ were selected for analysis.
The HIJING model emphasizes the effects of mini-jets non-flow correlations, while the AMPT model is used to investigate the flow angle fluctuations.
The AMPT model, which is employed to study the physical process in  relativistic heavy-ion collisions~\cite{Lin:2004en,Ma:2016fve,Ma:2013gga,Ma:2013uqa,Bzdak:2014dia,Nie:2018xog,Haque:2019vgi,Zhao:2019kyk,Bhaduri:2010wi,Nasim:2010hw,Xu:2010du,Magdy:2020bhd,Guo:2019joy,Magdy:2020gxf,Magdy:2022cvt}. The AMPT model includes several important model elements: (i) an initial partonic state provided by the HIJING model~\cite{Wang:1991hta,Gyulassy:1994ew}, with the HIJING model parameters, 
$a=0.55$ and $b=0.15$ GeV$^{-2}$ are used for the Lund string fragmentation function 
$f(z) \propto z^{-1} (1-z)^a
\exp (-b~m_{\perp}^2/z)$, 
where $z$ represents the light-cone momentum fraction of the yielded hadron of transverse mass $m_\perp$ about that of the fragmenting string. Also, (ii) partonic scattering with a cross-section,
\begin{eqnarray}\label{eq:21}
\sigma_{pp} &=& \dfrac{9 \pi \alpha^{2}_{s}}{2 \mu^{2}},
\end{eqnarray}
where $\mu$ is the screening mass and $\alpha_{s}$  represents the coupling constant of the QCD. {  The partonic scattering with cross-section $\sigma_{pp} $ drives the expansion dynamics~\cite{Zhang:1997ej}; }
(iii) the hadronization process through coalescence followed by the hadronic interactions~\cite{Li:1995pra}. 
In the current work, the $\alpha_{s}$ and  $\mu$ are fixed to 0.47 and 3.41~$fm^{-1}$~\cite{Xu:2011fi} respectively.
%--------------------------------------------------------------------

%--------------------------------------------------------------------
\subsection{Analysis Method}
%--------------------------------------------------------------------

%--------------------------------------------------------------------
 The events generated were analyzed using the  two- and multi-particle correlations given via the use of the subevents cumulant methods~\cite{Jia:2017hbm,Huo:2017nms,Zhang:2018lls,Magdy:2020bhd}. The observables discussed in this work can be given in terms of the flow vectors as;
%--------------------------------------------------------------------
\begin{eqnarray}\label{eq:2-1}
    Q_{n,X_j} &=& \sum_{i} e^{\mathit{i} n \phi_{i} },
\end{eqnarray}
where $\phi_i$ is the azimuthal angle of the $\mathit{i}^{th}$ particle in the $X_j$  subevent.
%--------------------------------------------------------------------

%--------------------------------------------------------------------
\begin{figure}[!h] 
\includegraphics[width=0.99 \linewidth, angle=-0,keepaspectratio=true,clip=true]{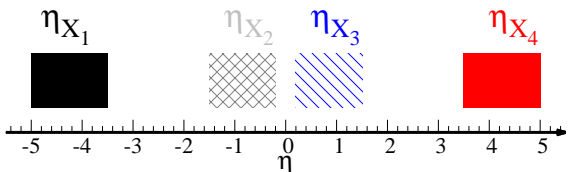}
\vskip -0.4cm
\caption{The $\eta$ ranges for the subevents used in this work.\label{fig:0}}
\vskip -0.1cm
\end{figure}
%--------------------------------------------------------------------
%--------------------------------------------------------------------
In Fig.~\ref{fig:0}, { events are partitioned into four subevents,  $\eta_{X_1}$ and $\eta_{X_4}$ as an integrated flow vector and $\eta_{X_2}$ and $\eta_{X_3}$ as a differential flow vector}.
The current work contains two assumptions; first the two integrated(differential) event planes are equivalent (i.e., $\psi_{X_1,n}\sim\psi_{X_4,n}\sim\psi_{n}$ and $\psi^{d}_{X_2}~\sim~\psi^{d}_{X_3}~\sim~\psi_{d,n}$). Second, non-flow effects are very small for the two-particle correlations with a large $\eta$ gap. %The second  assumptions is validated in this work using the HIJING model.

%--------------------------------------------------------------------
The two-particle correlations can be given as;
%--------------------------------------------------------------------
\begin{eqnarray}\label{eq:2-2}
    \langle 2 \rangle_{X_1 X_1} &=& e^{\mathit{i} n (\phi_{1,X_1} - \phi_{2,X_1}) } \equiv \dfrac{ Q_{n,X_1}Q^{*}_{n,X_1} - M_{X_1}}{M_{X_1} (M_{X_1}-1)},\\  \nonumber
    \langle 2 \rangle_{X_1 X_3} &=& e^{\mathit{i} n (\phi_{1,X_1} - \phi_{2,X_3}) } \equiv  \dfrac{ Q_{n,X_1}Q^{*}_{n,X_3}}{M_{X_1} M_{X_3}},
\end{eqnarray}
%--------------------------------------------------------------------
where $M_{X}$ is the number of particles in subevent $X$.
%--------------------------------------------------------------------

The overall two-particle correlations can be rewritten as;
%--------------------------------------------------------------------
\begin{eqnarray}\label{eq:2-3}
 \langle \langle 2 \rangle \rangle &\sim & \langle \langle 2 \rangle \rangle_{g} + \delta_{2} 
\end{eqnarray}
%--------------------------------------------------------------------
where $ \langle \langle ~ \rangle \rangle$ refer to average over all tracks and average over all events, $\delta_{2} \thicksim \delta_{2,short} + \delta_{2,long}$ is non-flow from two-particle correlations and {  $\langle \langle 2 \rangle \rangle_{g}$ is the two-particle flow correlations (i.e., determined by the geometry of the collision system)}. Note that short-range non-flow can be reduced using the subevents methods with $\eta$ gap~\cite{Magdy:2021sba}.

%--------------------------------------------------------------------
The four-particle correlations can be given using the two-subevents as;
{\small
\begin{eqnarray}\label{eq:2-4}
    \langle 4 \rangle_{T_1, X_1 X_3} &=& e^{\mathit{i} n (\phi_{1,X_1} + \phi_{1,X_1} - \phi_{3,X_3} - \phi_{4,X_3}) } \\  \nonumber
    &\equiv&  \dfrac{ (Q^{2}_{n,X_1} - Q_{2n,X_1})  (Q^{2}_{n,X_3} - Q_{2n,X_3})^{*}}{M_{X_1} (M_{X_1}-1)  M_{X_3} (M_{X_3}-1)},
\end{eqnarray}
}
{\small
\begin{eqnarray}\label{eq:2-5}
    \langle 4 \rangle_{T_2, X_1 X_3} &=& e^{\mathit{i} n (\phi_{1,X_1} + \phi_{1,X_3} - \phi_{3,X_1} - \phi_{4,X_3}) } \\  \nonumber
    &\equiv&  \dfrac{ (Q^{2}_{n,X_1} - M_{X_1})  (Q^{2}_{n,X_3} - M_{X_3})^{*}}{M_{X_1} (M_{X_1}-1)  M_{X_3} (M_{X_3}-1)},
\end{eqnarray}
}
where $T_{1,2}$ represent type $I$, and $II$ for the four-particle correlations.
%--------------------------------------------------------------------

%--------------------------------------------------------------------
The four-particle correlations can be given using the three-subevents as;
{\small
\begin{eqnarray}\label{eq:2-6}
    \langle 4 \rangle_{T_1, X_1 X_2 X_3} &=& e^{\mathit{i} n (\phi_{1,X_1} + \phi_{1,X_1} - \phi_{3,X_2} - \phi_{4,X_3}) } \\  \nonumber
    &\equiv&  \dfrac{ (Q^{2}_{n,X_1} - Q_{2n,X_1})  (Q_{n,X_2}  Q_{n,X_3})^{*}}{M_{X_1} (M_{X_1}-1)  M_{X_2} M_{X_3} },
\end{eqnarray}
}
{\small
\begin{eqnarray}\label{eq:2-7}
    \langle 4 \rangle_{T_2, X_1 X_2 X_3} &=& e^{\mathit{i} n (\phi_{1,X_1} + \phi_{1,X_2} - \phi_{3,X_1} - \phi_{4,X_3}) } \\  \nonumber
    &\equiv&  \dfrac{ (Q^{2}_{n,X_1} - M_{X_1})  Q^{}_{n,X_2} Q^{*}_{n,X_3}}{M_{X_1} (M_{X_1}-1)  M_{X_2} M_{X_3}}.
\end{eqnarray}
}
%--------------------------------------------------------------------

%--------------------------------------------------------------------
The four-particle correlations can be given using the four-subevents as;
{\small
\begin{eqnarray}\label{eq:2-8}
    \langle 4 \rangle_{T_1, X_1 X_2 X_3 X_4} &=& e^{\mathit{i} n (\phi_{1,X_1} + \phi_{1,X_2} - \phi_{3,X_3} - \phi_{4,X_4}) } \\  \nonumber
    &\equiv&  \dfrac{ (Q_{n,X_1} Q^{*}_{n,X_3})  (Q_{n,X_2}  Q^{*}_{n,X_4}) }{M_{X_1} M_{X_2} M_{X_3} M_{X_4} },
\end{eqnarray}
}
{\small
\begin{eqnarray}\label{eq:2-8x}
    \langle 4 \rangle_{T_2, X_1 X_2 X_3 X_4} &=& e^{\mathit{i} n (\phi_{1,X_1} + \phi_{1,X_2}  - \phi_{4,X_4} - \phi_{3,X_3}) } \\  \nonumber
    &\equiv&  \dfrac{ (Q_{n,X_1} Q^{*}_{n,X_4})  (Q_{n,X_2}  Q^{*}_{n,X_3}) }{M_{X_1} M_{X_2} M_{X_3} M_{X_4} }.
\end{eqnarray}
}
%--------------------------------------------------------------------
Note that the integrated $\langle 4 \rangle_{T_1, X_1 X_2 X_3 X_4}$ and $\langle 4 \rangle_{T_2, X_1 X_2 X_3 X_4}$ are equivalent by construction.

%--------------------------------------------------------------------
In general the different contributions to the four-particle correlations~\cite{Jia:2017hbm} can be given as:
\begin{eqnarray}\label{eq:2-9}
 \langle \langle 4 \rangle \rangle &\sim& \langle \langle 4 \rangle \rangle_{g} + \delta_{4}  +   \delta_{2},
\end{eqnarray}
where $\delta_{4}$ is non-flow from four-particle correlations and $ \langle \langle 4 \rangle \rangle_{g}$ is the four-particle flow correlations.\\
%--------------------------------------------------------------------

%--------------------------------------------------------------------
Note that using type-I two-, three- and four-subevets with $\eta$ gaps, most of non-flow in the four-particle correlations is reduced.
In contrast, type-II will contain more non-flow~\cite{Jia:2017hbm}. The latter can be reduce by subtracting the two-particle correlations Eq.~\ref{eq:2-2} in the $k$-particle cumulants~\cite{Jia:2017hbm}. The four-particle cumulants are given as: 
%--------------------------------------------------------------------
%{\small
\begin{eqnarray}\label{eq:2-10}
  C_{T_1, X_1 X_3}       &=&  \langle \langle 4 \rangle \rangle_{T_1, X_1 X_3}     - 2 \langle \langle 2 \rangle \rangle_{X_1 X_3} \langle \langle 2 \rangle \rangle_{X_1 X_3}, \\ \nonumber
  C_{T_1, X_1 X_2 X_3}   &=&  \langle \langle 4 \rangle \rangle_{T_1, X_1 X_2 X_3} - 2 \langle \langle 2 \rangle \rangle_{X_1 X_2} \langle \langle 2 \rangle \rangle_{X_1 X_3}, \\ \nonumber
  C_{T_1, X_1 X_2 X_3 X_4}   &=&  \langle \langle 4 \rangle \rangle_{T_1, X_1 X_2 X_3 X_4} -  \langle \langle 2 \rangle \rangle_{X_1 X_3} \langle \langle 2 \rangle \rangle_{X_2 X_4}, \\ \nonumber
  &-&  \langle \langle 2 \rangle \rangle_{X_1 X_4} \langle \langle 2 \rangle \rangle_{X_1 X_3}
\end{eqnarray}
%}
%{\larg
\begin{eqnarray}\label{eq:2-11}
C_{T_2, X_1 X_3}       &=&  \langle \langle 4 \rangle \rangle_{T_2, X_1 X_3} -  \langle \langle 2 \rangle \rangle_{X_1 X_1} \langle \langle 2 \rangle \rangle_{X_3 X_3}  \\ \nonumber 
&-& \langle \langle 2 \rangle \rangle_{X_1 X_3} \langle \langle 2 \rangle \rangle_{X_1 X_3},  \\ \nonumber
C_{T_2, X_1 X_2 X_3}   &=&  \langle \langle 4 \rangle \rangle_{T_2, X_1 X_2 X_3} -  \langle \langle 2 \rangle \rangle_{X_1 X_1} \langle \langle 2 \rangle \rangle_{X_2 X_3}  \\ \nonumber 
&-& \langle \langle 2 \rangle \rangle_{X_1 X_2} \langle \langle 2 \rangle \rangle_{X_1 X_3}, \\ \nonumber 
C_{T_2, X_1 X_2 X_3 X_4}   &=&  \langle \langle 4 \rangle \rangle_{T_2, X_1 X_2 X_4 X_3} -  \langle \langle 2 \rangle \rangle_{X_1 X_3} \langle \langle 2 \rangle \rangle_{X_2 X_4}  \\ \nonumber 
&-& \langle \langle 2 \rangle \rangle_{X_1 X_4} \langle \langle 2 \rangle \rangle_{X_2 X_3}.
\end{eqnarray}
%}
%--------------------------------------------------------------------
Note that the integrated $C_{T_1, X_1 X_2 X_3 X_4}$ and $C_{T_2, X_1 X_2 X_3 X_4}$ are equivalent by construction.  \\

%--------------------------------------------------------------------
%--------------------------------------------------------------------
\subsubsection{The flow angle fluctuations} %--------------------------------------------------------------------

%--------------------------------------------------------------------
The differential/integrated two-particle correlations via the  two-subevents method with $\eta$ gap can be given as;
\begin{eqnarray}\label{eq:3-1a}
    \langle \langle 2 \rangle \rangle_{X_1 X^{d}_3} &\sim & \langle v_{n,X_1} v_{n,d,X_3} cos\left( n \left( \psi_{n, X_1} - \psi_{n, d, X_3} \right) \right) \rangle \nonumber  \\ 
    &\sim & \langle v_{d,n} v_{n} cos\left( n \left( \psi_{n,d} - \psi_{n} \right) \right) \rangle ,
\end{eqnarray}

%--------------------------------------------------------------------
\begin{eqnarray}\label{eq:3-1b}
    \langle \langle 2 \rangle \rangle_{X^{d}_1 X^{d}_3} &\sim & \langle v_{n,d,X_1} v_{n,d,X_3} cos\left( n \left( \psi_{n,d, X_1} - \psi_{n, d, X_3} \right) \right) \rangle \nonumber   \nonumber  \\
    &\sim & \langle v_{d,n} v_{d,n} \rangle.
\end{eqnarray}
%--------------------------------------------------------------------

%--------------------------------------------------------------------
Also the type-I differential four-particle correlations are given as;
{\small
\begin{eqnarray}\label{eq:3-2}
  \langle  \langle 4 \rangle \rangle_{T1, X_1 X^{d}_3}     &\sim & \langle v^{2}_{n,X_1} v^{2}_{n, d, X_3} \nonumber  \\
  & & cos\left( 2 n \left( \psi_{n,X_1} - \psi_{n, d, X_3} \right) \right) \rangle  \nonumber  \\
   &\sim & \langle v^{2}_{n} v^{2}_{n,d}  cos\left( 2 n \left( \psi_{n} - \psi_{n,d} \right) \right) \rangle,
\end{eqnarray}
%--------------------------------------------------------------------
\begin{eqnarray}\label{eq:3-3}   
  \langle  \langle 4 \rangle \rangle_{T1, X_1 X^{d}_2 X^{d}_3} &\sim & \langle v^{2}_{n,X_1} v_{n,d, X_2} v_{n,d, X_3}   \nonumber \\ 
  & &  cos\left( n \left( \psi_{n,X_1} - \psi_{n,d, X_2} \right) \right)  \nonumber \\ 
  & &  cos\left( n \left( \psi_{n,X_1} - \psi_{n,d, X_3} \right) \right) \rangle  \nonumber \\ 
  &\sim & \langle v^{2}_{n} v^{2}_{n,d}  cos\left( 2 n \left( \psi_{n} - \psi_{n,d} \right) \right) \rangle,
\end{eqnarray}
%--------------------------------------------------------------------
\begin{eqnarray}\label{eq:3-4}   
  \langle  \langle 4 \rangle \rangle_{T1, X_1 X^{d}_2 X^{d}_3 X_4} &\sim & \langle v_{n,X_1} v_{n,d,X_2} v_{n,d,X_3} v_{n,X_4}  \nonumber \\ 
  & &  cos\left( n \left( \psi_{n,X_1} - \psi_{n,d,X_3} \right) \right)  \nonumber  \\
  & &  cos\left( n \left( \psi_{n,d,X_2} - \psi_{n,X_4} \right) \right) \rangle  \nonumber \\
    &\sim & \langle v^{2}_{n} v^{2}_{n,d}  cos\left( 2 n \left( \psi_{n} - \psi_{n,d} \right) \right) \rangle.
\end{eqnarray}
}
%--------------------------------------------------------------------

%--------------------------------------------------------------------
The type-II differential four-particle correlations are given as;
{\small
\begin{eqnarray}\label{eq:3-5}
  \langle  \langle 4 \rangle \rangle_{T2, X_1 X^{d}_3}     &\sim & \langle v^{2}_{n,X_1} v^{2}_{n, d, X_3} \nonumber  \\
  & & cos\left( n \left( \psi_{n,X_1} - \psi_{n,X_1} \right) \right) \rangle  \nonumber  \\
  & & cos\left( n \left( \psi_{n, d, X_3} - \psi_{n, d, X_3} \right) \right) \rangle  \nonumber  \\ 
  & + & \delta_{2,X_1} \delta_{2,X_3}\nonumber \\
  &\sim & \langle v^{2}_{n} v^{2}_{n,d}  \rangle + \delta_{2,X_1} \delta_{2,X_3},
\end{eqnarray}
%--------------------------------------------------------------------
\begin{eqnarray}\label{eq:3-6}   
  \langle  \langle 4 \rangle \rangle_{T2, X_1 X^{d}_2 X^{d}_3} &\sim & \langle v^{2}_{n,X_1} v_{n,d, X_2} v_{n,d, X_3}   \nonumber \\ 
  & & cos\left( n \left( \psi_{n,X_1} - \psi_{n,X_1} \right) \right)  \nonumber \\ 
  & & cos\left( n \left( \psi_{n,d, X_2} - \psi_{n,d, X_3} \right) \right) \rangle  \nonumber \\ 
  &+& \delta_{2,X_1} \nonumber \\
  &\sim & \langle v^{2}_{n} v^{2}_{n,d} \rangle  + \delta_{2,X_1},
\end{eqnarray}
%--------------------------------------------------------------------
\begin{eqnarray}\label{eq:3-7}   
  \langle  \langle 4 \rangle \rangle_{T2, X_1 X^{d}_2 X^{d}_3 X_4} &\sim & \langle v_{n,X_1} v_{n,d,X_2} v_{n,d,X_3} v_{n,X_4}  \nonumber \\ 
  & &  cos\left( n \left( \psi_{n,X_1}   - \psi_{n,X_4}   \right) \right)  \nonumber  \\
  & &  cos\left( n \left( \psi_{n,d,X_2} - \psi_{n,d,X_3} \right) \right)  \rangle     \nonumber  \\ 
  &\sim & \langle v^{2}_{n} v^{2}_{n,d} \rangle.
\end{eqnarray}
%--------------------------------------------------------------------

%--------------------------------------------------------------------
\begin{itemize}
%--------------------------------------------------------------------
\item{ The differential flow angle fluctuations using the two-particles correlations and two-subevents can be given as:}
\begin{eqnarray}\label{eq:3-8}
  r_{2}\lbrace 2,2S\rbrace(p_T) &=&  \dfrac{\langle \langle 2 \rangle \rangle_{X_1 X^{d}_3}}{\sqrt{\langle \langle 2 \rangle \rangle_{X^{d}_1 X^{d}_3}  \langle \langle 2 \rangle \rangle_{X_1 X_3} }}  \nonumber  \\
              &\sim &  cos\left( n \left( \psi_{n,d} - \psi_{n} \right) \right)
\end{eqnarray}
%--------------------------------------------------------------------

%--------------------------------------------------------------------
\item{ The differential flow angle fluctuations using the four-particles and two-subevents can be given as:}
\begin{eqnarray}\label{eq:3-9}
   r_{2}\lbrace 4,2S\rbrace(p_T) &=&  \dfrac{\langle \langle 4 \rangle \rangle_{T_1, X_1 X^{d}_3}}{\langle \langle 4 \rangle \rangle_{T_2, X_1 X^{d}_3}}  \nonumber  \\
              &\sim &  \dfrac{\langle v^{2}_{n} v^{2}_{n,d}  cos\left( 2 n \left( \psi_{n} - \psi_{n,d} \right) \right) \rangle}{\langle v^{2}_{n} v^{2}_{n,d}\rangle + \delta_{2, X_1} \delta_{2, X_3}  }.
\end{eqnarray}
%--------------------------------------------------------------------

%--------------------------------------------------------------------
\item{ The differential flow angle fluctuations using the four-particle and three-subevents can be given as:}
\begin{eqnarray}\label{eq:3-10}
   r_{2}\lbrace 4,3S\rbrace(p_T) &=&  \dfrac{\langle \langle 4 \rangle \rangle_{T_1, X_1 X^{d}_2 X^{d}_3}}{\langle \langle 4 \rangle \rangle_{T_2, X_1 X^{d}_2  X^{d}_3}}  \nonumber  \\
              &\sim &  \dfrac{\langle v^{2}_{n} v^{2}_{n,d}  cos\left( 2 n \left( \psi_{n} - \psi_{n,d} \right) \right) \rangle}{\langle v^{2}_{n} v^{2}_{n,d}\rangle + \delta_{2, X_1}}.
\end{eqnarray}
%--------------------------------------------------------------------

%--------------------------------------------------------------------
\item{ The differential flow angle fluctuations using the four-particle and four-subevents can be given as:}
\begin{eqnarray}\label{eq:3-11}
   r_{2}\lbrace 4,4S\rbrace(p_T) &=&  \dfrac{\langle \langle 4 \rangle \rangle_{T_1, X_1 X^{d}_2 X^{d}_3 X_4}}{\langle \langle 4 \rangle \rangle_{T_2, X_1 X^{d}_2  X^{d}_3 X_4}}  \nonumber  \\
              &\sim &  \dfrac{\langle v^{2}_{n} v^{2}_{n,d}  cos\left( 2 n \left( \psi_{n} - \psi_{n,d} \right) \right) \rangle}{\langle v^{2}_{n} v^{2}_{n,d}\rangle}   \nonumber  \\
               &\sim &  \langle cos\left( 2 n \left( \psi_{n} - \psi_{n,d} \right) \right) \rangle
\end{eqnarray}
%--------------------------------------------------------------------

%--------------------------------------------------------------------
\end{itemize}
%--------------------------------------------------------------------

%--------------------------------------------------------------------
%--------------------------------------------------------------------
\section{Results and discussion}\label{sec:3}
%--------------------------------------------------------------------
The reliability of the extracted flow angles $\Psi_n$ fluctuations can be influenced by possible short- and long-range non-flow contributions to the two- and four-particle correlators used for the extractions. Therefore, it is informative to consider a figure of merit for these contributions to the four-particle correlations.
%--------------------------------------------------------------------
%--------------------------------------------------------------------
\begin{figure}[!h] 
\includegraphics[width=1.0  \linewidth, angle=-0,keepaspectratio=true,clip=true]{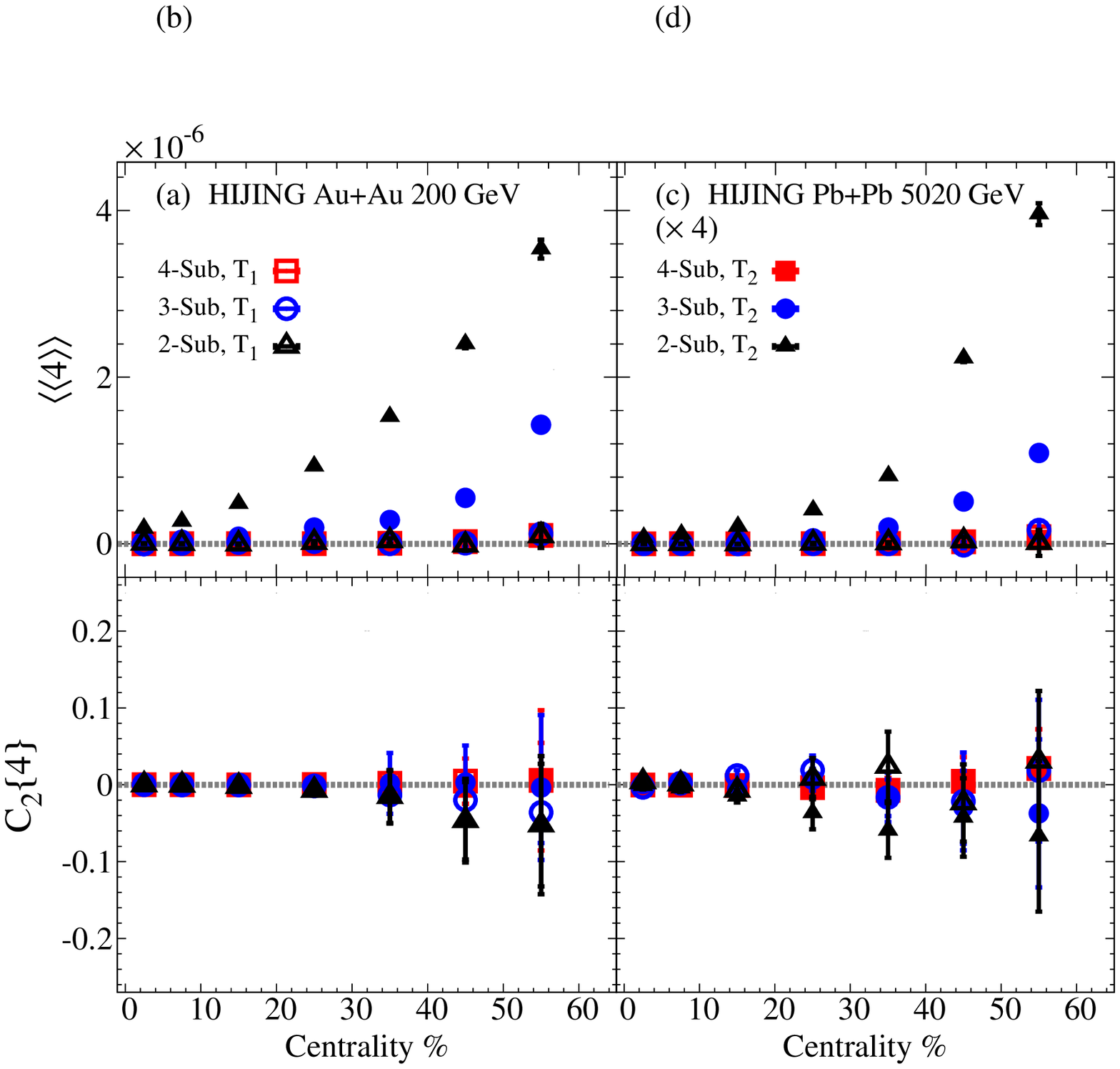}
\vskip -0.4cm
\caption{The centrality dependence of the four-particle correlations panels and the four-particle cumulant for Au+Au at $\sqrt{\textit{s}_{NN}}~=$ 200~GeV panels (a) and (b) and for Pb+Pb at $\sqrt{\textit{s}_{NN}}~=$ 5020~GeV from the HIJING model using the two-, three-, and four-subevents correlation methods.
}\label{fig:1}
\vskip -0.3cm
\end{figure}
%--------------------------------------------------------------------
%--------------------------------------------------------------------
\begin{figure}[!h] 
\includegraphics[width=1.0 \linewidth, angle=-0,keepaspectratio=true,clip=true]{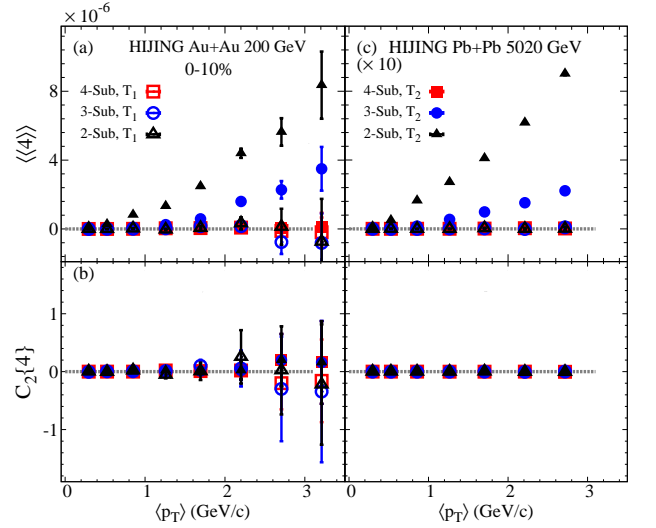}
\vskip -0.4cm
\caption{The 0-10\% central $p_T$ dependence of the four-particle correlations panels and the four-particle cumulant for Au+Au at $\sqrt{\textit{s}_{NN}}~=$ 200~GeV panels (a) and (b) and for Pb+Pb at $\sqrt{\textit{s}_{NN}}~=$  5020~GeV from the HIJING model using the two-, three-, and four-subevents correlation methods.
}\label{fig:2}
\vskip -0.3cm
\end{figure}
%--------------------------------------------------------------------
%--------------------------------------------------------------------

%--------------------------------------------------------------------
%--------------------------------------------------------------------
The non-flow effects on the  (I) the four-particle correlations ($\langle \langle 4 \rangle \rangle$) and (II) four-particle cumulant ($C_{2}\lbrace 4 \rbrace$) are shown in Figs.~\ref{fig:1} and~\ref{fig:2}.  The results are presented for Au+Au at $\sqrt{\textit{s}_{NN}}~=$ 200~GeV and Pb+Pb at $\sqrt{\textit{s}_{NN}}~=$ 5020~GeV from the HIJING model using the two-, three-, and four-subevents methods.
%--------------------------------------------------------------------
%
%--------------------------------------------------------------------
The HIJING model results in Fig.~\ref{fig:1} panels (a) and (c) show the centrality dependence of the $\langle \langle 4 \rangle \rangle$. The type-I (all subevents) and type-II (four-subevents) $\langle \langle 4 \rangle \rangle$ are much reduced, indicating little if any non-flow effects. In contrast, the two- and the three-subevents type-II $\langle \langle 4 \rangle \rangle$ show a strong centrality dependence indicating non-flow effects from the $\langle \langle 2 \rangle \rangle$ as pointed out in Eqs.~\ref{eq:2-9}, \ref{eq:3-5}, and~\ref{eq:3-6}. The difference in values between the two presented energies is expected to be related to the multiplicity difference at the same centrality. 
In addition, the $C_{2}\lbrace 4 \rbrace$, which subtracts the two-particle correlations ( panels (b) and (d)) are shown to be zero indicating their ability to reduce non-flow effects. These results are compatible with other investigations~\cite{Huo:2017nms,Zhang:2021phk}.
%--------------------------------------------------------------------

%--------------------------------------------------------------------
The $\langle p_T \rangle$ dependence of the $\langle \langle 4 \rangle \rangle$ for 0-10\% central collisions are given in Fig.~\ref{fig:2} (panels (a) and (c)). My results show that 
 type-I (all subevents) and type-II (four-subevents) $\langle \langle 4 \rangle \rangle$ are much reduced, indicating small non-flow effects. The two- and the three-subevents type-II $\langle \langle 4 \rangle \rangle$ show a strong $\langle p_T \rangle$ dependence {  indicating that non-flow effects in the HIJING model are $\langle p_T \rangle$ dependent.}
As pointed out in Fig.~\ref{fig:1} the $C_{2}\lbrace 4 \rbrace$ which subtracts the two-particle correlations (panels (b) and (d)) show little if any non-flow correlations.
%--------------------------------------------------------------------

%--------------------------------------------------------------------
\begin{figure}[!h] 
\includegraphics[width=0.9 \linewidth, angle=-0,keepaspectratio=true,clip=true]{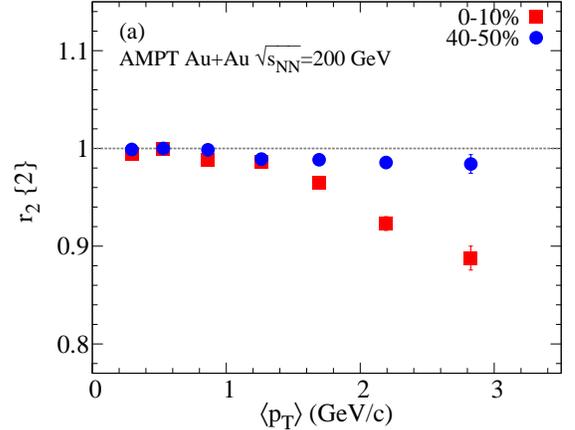}
\vskip -0.4cm
\caption{
The $\langle p_T \rangle$ dependence of the $r_{2}\lbrace 2 \rbrace$($p_T$) correlator for 0-10\% and 40-50\% central collisions for Au+Au at $\sqrt{\textit{s}_{NN}}~=$  200~GeV from the AMPT model.
}\label{fig:3}
\vskip -0.3cm
\end{figure}
%--------------------------------------------------------------------
%--------------------------------------------------------------------
%--------------------------------------------------------------------
%--------------------------------------------------------------------
\begin{figure}[!h] 
\includegraphics[width=1.0 \linewidth, angle=-0,keepaspectratio=true,clip=true]{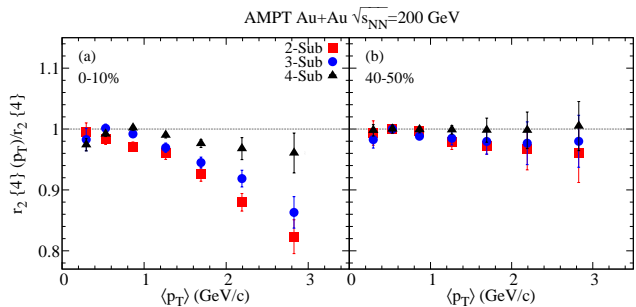}
\vskip -0.4cm
\caption{
The $\langle p_T \rangle$ dependence of the normalized $r_{2}\lbrace 4 \rbrace$($p_T$) correlator using the two-, three-, and four-subevents for 0-10\% (a) and 40-50\% (b) central collisions for Au+Au at $\sqrt{\textit{s}_{NN}}~=$ 200~GeV from the AMPT model.
}\label{fig:4}
\vskip -0.3cm
\end{figure}
%--------------------------------------------------------------------
%--------------------------------------------------------------------

%--------------------------------------------------------------------
{ 
In this work, the HIJING model is used as a testing ground for non-flow effects on the $\langle \langle 4 \rangle \rangle$ that are an essential part of $r_{2}\lbrace 4 \rbrace$. 
The results show that the two- and three-subevents results are contaminated by non-flow effects, which increase with centrality and $\langle p_T \rangle$. Such an effect could have an impact on the recent ALICE measurements~\cite{ALICE:2022smy} which use the two-subevents method with a small $\eta$ gap.
In contrast, the HIJING model results in Figs.~\ref{fig:1}  and~\ref{fig:2} suggest that the four-subevents cumulant method reduces the non-flow effects to less than 0.5\%, albeit model dependent. Experimental measurements of comparable magnitude would, of course, be challenging to interpret.
Although the four-subevents cumulant method is statistically demanding and often requires a wide $\eta$ acceptance, such a study can be achieved in the STAR experiment at RHIC~\cite{Llope:1997et}  via the the STAR Event Plane Detector (EPD)~\cite{Adams:2019fpo}.
}

Figure~\ref{fig:3} compares the $\langle p_T \rangle$ dependence of the two-particle 
$r_{2}\lbrace 2 \rbrace$ for 0-10\% and 40-50\% central collisions for Au+Au at $\sqrt{\textit{s}_{NN}}~=$  200~GeV from the AMPT model. The results indicated large flow angle fluctuations in central collisions with increasing the $\langle p_T \rangle$. Such an effect is largely reduced in peripheral collisions. My results are compatible with the prior calculations in Ref~\cite{Bozek:2021mov}. The $r_{2}\lbrace 2 \rbrace$ is expected to have remaining long-range non-flow effects.
%--------------------------------------------------------------------

%--------------------------------------------------------------------
The $\langle p_T \rangle$ dependence of the four-particle $r_{2}\lbrace 4 \rbrace$ using the two-, three-, and four-subevents method for 0-10\% (a) and 40-50\% (b) central collisions for Au+Au at $\sqrt{\textit{s}_{NN}}~=$ 200~GeV from the AMPT model shown in Fig.~\ref{fig:4}. In central collisions $r_{2}\lbrace 4 \rbrace$ indicated a $\langle p_T \rangle$ dependence with about 20\%--15\% variation when using  two-, and three-subevents method.
Using the four-subevents method gives an  $r_{2}\lbrace 4 \rbrace$ with about 4\% variation, which reflects the power of the such method to reduce non-flow effects. Such effects are vastly reduced in peripheral collisions, as shown in panel (b).
These calculations, can be done using the EPD~\cite{Adams:2019fpo} of the STAR experiment at RHIC~\cite{Llope:1997et}.
%--------------------------------------------------------------------

%--------------------------------------------------------------------
\section{Summary}\label{sec:4}
%--------------------------------------------------------------------
Multi-particle azimuthal correlations between different subevents have been used to study the nature of the flow angle fluctuations in Au+Au and Pb+Pb at 200 and 5020~GeV, respectively. 
 Using the HIJING model that contains only non-flow effects, I show that the flow angle fluctuations measurements using two- and three-subevents are likely contaminated by non-flow effects.
 By constructing an azimuthal correlation between four pseudorapidity ranges, I showed that the calculations of $r_{2}\lbrace 4 \rbrace$ are much less susceptible to these sources of non-flow. 
 In addition, using the AMPT model for Au+Au at 200 GeV, I predicted the $\langle p_T \rangle$ dependence of $r_{2}\lbrace 4 \rbrace$ using the four-subevents correlation method. 
 These studies suggest that the measurements of the $r_{2}\lbrace 4 \rbrace$ need to be measured with the four-subevents methods before any physics conclusion can be made.
%--------------------------------------------------------------------

%--------------------------------------------------------------------
\section*{Acknowledgments}
The author thanks Emily Racow, J. Jia, C. Zhang and P. Bozek for the useful discussions and for pointing out important references.
This research is supported by the US Department of Energy, Office of Nuclear Physics (DOE NP),  under contracts DE-FG02-87ER40331.A008.
%--------------------------------------------------------------------

%--------------------------------------------------------------------
%\bibliographystyle{aipauth4-1}
\bibliography{ref} 
%--------------------------------------------------------------------
\end{document}